\documentclass[aps,pra,twocolumn,showkeys,showpacs,a4paper,nobibnotes,superscriptaddress]{revtex4-2}

\usepackage{graphicx}
\usepackage{amsmath}
\usepackage[colorlinks,citecolor=blue,linkcolor=blue,urlcolor=blue]{hyperref}
\usepackage{soul}

\usepackage{multirow}
\usepackage{booktabs}
\usepackage{tabularx}

\begin{document}

\title{Electron capture induced fragmentation of CO$_2^{3+}$: Influence of projectile charge on sequential and concerted break-up pathways}

\author{Akash Srivastav}
\email[]{akash.srivastav@students.iiserpune.ac.in}
\affiliation{Indian Institute of Science Education and Research Pune, Homi Bhabha Road, Pune~411008, India}
\author{Sumit Srivastav}
\affiliation{Indian Institute of Science Education and Research Pune, Homi Bhabha Road, Pune~411008, India}
\affiliation{\textup{Present address:} Normandie Univ., ENSICAEN, UNICAEN, CEA, CNRS, CIMAP, 14000 Caen, France}
\author{Bhas Bapat}
\affiliation{Indian Institute of Science Education and Research Pune, Homi Bhabha Road, Pune~411008, India}


\begin{abstract}
	
We investigate the \mbox{O$^+$:\,C$^+$:\,O$^+$} fragmentation channel of CO$_2^{3+}$ produced in slow collisions with Ar$^{q+}$ projectiles ($4 \le q \le 16$, velocities $\approx 0.3$~a.u). Using the native-frames method, we disentangle the sequential and concerted break-up processes and their corresponding kinetic energy release (KER) distributions. \textit{Ab initio} potential energy curves of CO$_2^{3+}$ are calculated and mapped to the KER spectra to identify the underlying electronic states involved in the fragmentation. While the sequential KER distributions remain nearly unchanged for across the projectile charge range, the concerted KER distributions exhibit pronounced but non-systematic variations with projectile charge. In addition, a low KER feature around 15.5~eV---previously associated with sequential break-up in electron and proton impact---is observed for Ar$^{4+}$ impact and, to a lesser extent, for Ar$^{6+}$ impact. It originates predominantly from concerted break-up of the low-lying $^2\Pi_\text{g}$ and $^{2,4}\Pi_\text{u}$ states. Branching ratios of the two break-up pathways deviate from simple monotonic trends for certain projectiles, but barring these exceptions, the fraction of concerted break-up decreases with increasing $q$, while that for sequential break-up increases. These findings underscore the necessity of accounting for the detailed electronic structure of the projectile, rather than its charge alone, to achieve a comprehensive understanding of collisional dynamics in slow, highly charged ion collisions.

\end{abstract}


\maketitle

\section{INTRODUCTION}

The study of the formation and fragmentation dynamics of multiply charged molecular ions (MCMI) represent a central and rapidly evolving area of research in contemporary atomic and molecular physics. One prominent route to produce these species is via collisions with highly charged ions (HCI), a process that has broad relevance in fundamental physics, planetary science \cite{Dennerl2010,refId,Kharchenko_2006,Larsson_2012}, and radiation biology \cite{Amaldi_2005,RevModPhys.82.383}. Depending on the projectile velocity ($v$), the creation of MCMI proceeds via mechanisms such as direct ionization in which all ionized electrons are ejected into the continuum, multielectron capture, or transfer ionization (TI) where some electrons are captured and others are ejected to the continuum. For slow collisions ($v<1$~a.u.) multielectron capture is the dominant process \cite{Vancura_1994,Wu_1995,Wells_2005,Luna_2016}. The capture processes are particularly pivotal in slow HCI-neutral collisions \cite{JANEV1985265,M_Barat_1992} and bear significant importance in astrophysics \cite{Dennerl2010,Cravens_2001,Snowden_2004,refId,Kharchenko_2006} as well as plasma diagnostics in fusion research \cite{R_C_Isler_1994,10.1063/1.5132936}. 

Typically, MCMI are unstable and dissociate into charged and neutral fragments. The fragmentation may occur through a concerted process, where all bonds break nearly simultaneously, or through a sequential pathway, involving stepwise bond breaks via metastable intermediates.

Neumann \emph{et al.} \cite{Neumann_2010} showed that the total energy deposited into the target is the key parameter that controls the dynamics of sequential and concerted fragmentation. The deposited energy is statistically partitioned among various excited electronic states, and the relative populations of these states ultimately govern the dynamics of sequential and concerted break-up. 

From a classical viewpoint, the energy transferred to the target depends on the projectile's charge $q$, velocity $v$, and impact parameter $b$. Of these, $q$ and $v$ can be controlled in an experiment, and variations in these parameters are expected to alter the relative populations of various excited states, thereby influencing the overall fragmentation dynamics. Numerous studies have examined how changes in $q$ and $v$ affect dissociative ionization, particularly for diatomic and triatomic molecules. The sum of the kinetic energies of the fragments, or the kinetic energy release (KER), is a key parameter for understanding energy deposition and the consequent population of electronic states of the molecular ion. Hence it is important to experimentally investigate the dependence of KER on $q$ or $v$.

For diatomic molecules where only concerted break-up can occur, Folkerts \emph{et al.} studied CO fragmentation in collisions with 0.4~a.u.\ He$^{2+}$ and O$^{7+}$ ions \cite{Folkerts_1996}, concluding from KER distributions that lower $q$ leads to higher excitation of CO$^{2+}$. Khan \emph{et al.} explored $q$ and $v$ effects in O$^{q+}$---N$_2$ ($q=2,4,6$) system at $v=1.0$, $1.5$, and $2.0$~a.u.\ \cite{Khan_2021}, observing that increasing $q$ or decreasing $v$ consistently yields gentler collisions with lower energy transfer. For triatomic molecules, where concerted as well as sequential break-up can occur, Wales \emph{et al.} examined three-body break-up of OCS under 0.25~a.u.\ Ar$^{4+}$ impact and 0.35~a.u.\ Ar$^{8+}$ impact \cite{Wales_2012}, finding nearly identical KER distributions for $q=4,8$. For 0.49~a.u.\ Ar$^{q+}$ impact on CO$_2$ ($q=3,6,8$) collisions, Kumar \emph{et al.} found that Ar$^{6+}$ results in the widest KER distribution \cite{Kumar_2024}. Recent work by Srivastav \emph{et al.}, from this group, analyzed $q$ and $v$ dependence in CO$_2$ three-body break-up induced by protons and Ar$^{q+}$ ($8\leq q\leq 14$) projectiles \cite{Sumit_2021,Sumit_2022}.  They showed that small velocity changes of the proton significantly alter the break-up and that there is a pronounced dependence of the KER distribution on $q$.

Since both $q$ and $v$ influence the energy deposited in a collision, their effect will be reflected in the relative dominance of sequential and concerted break-up mechanisms. Although a number of studies have explored the dynamics of sequential and concerted fragmentation \cite{Neumann_2010,Wales_2012,Sumit_2021,Sumit_2022,Khan_2015,Wang_2015,Yan_2016,Rajput_2018,Severt_2024,Sandeep_2025}, only a few have examined their dependence on collisional parameters $q$ or $v$.  Where this has been studied, it is  within limited range of parameters \cite{Sumit_2021,Sumit_2022}. Therefore, an exploration of how sequential and concerted break-up dynamics change over a wide range of $q$ and $v$ is warranted.

In cases where a systematic trend in the KER distributions as a function of $q$ was observed at a fixed collision velocity (for $v\le 1.0$~a.u.), it was suggested that increasing the projectile charge leads to a gentler collision \cite{Folkerts_1996,Khan_2021}. However, recent studies suggest that this does not always hold, particularly in collision regimes dominated by multielectron capture \cite{Sumit_2021,Kumar_2024}. Within the quasi-molecular curve-crossing framework, appropriate for slow collisions, electron capture is treated as a quasi-resonant transition between entrance and exit channels. Here, electron capture depends sensitively on the binding energies of available final states in the post-collision projectile. Thus, the detailed electronic structure of the projectile, rather than its charge alone, may govern fragmentation dynamics.

In this work, we investigate the \mbox{O$^+$:\,C$^+$:\,O$^+$} fragmentation channel of CO$_2^{3+}$ produced by Ar$^{q+}$ ($4 \le q \le 16$) projectiles at a collision velocity of 0.27~a.u.\ for Ar$^{4+}$ and 0.31~a.u.\ for Ar$^{q+}$ ($q\ge6$). While this fragmentation channel has been widely studied using various projectiles \cite{Neumann_2010,Sumit_2022,Khan_2015,Wang_2015,Yan_2016,Kumar_2024,Sanderson_1999,Siegmann_2002,Jana_2011,Bhatt_2012}, our system closely resembles that of \cite{Sanderson_1999, Neumann_2010,Kumar_2024} but at a slightly lower velocity and over a broader projectile charge range. Applying the native-frames method \cite{Severt_2024,Rajput_2018}, we disentangle sequential and concerted fragmentation pathways and report their respective KER distributions. The KER distributions for sequential break-up are found to be almost similar for all projectiles, whereas those for concerted break-up exhibit pronounced variations but lack systematic dependence on $q$. We have computed \textit{ab initio} potential energy curves for various electronic states of CO$_2^{3+}$ and mapped them to the KER data to identify the participating electronic states. Furthermore, we examine the trends of the branching ratios of the two break-up pathways as a function of $q$.  Our findings show that to understand the collisional dynamics in these slow collision regimes it is important to take into account the electronic structure of the projectile and not just its charge.

\section{EXPERIMENTAL DETAILS}

Experiments were carried out with the electron beam ion source (EBIS) facility at IISER Pune. The EBIS is capable of delivering projectile ions in the energy range of (5$-$30) keV/$q$ \cite{Bapat_2020}. A multihit capable ion momentum spectrometer (IMS) satisfying the Wiley-McLaren space focusing condition \cite{Wiley_1955} was employed to investigate the fragmentation dynamics of molecular ions. The post-collision charge state analysis of the projectile can be done with a cylindrical deflector analyzer (CDA) coupled downstream of the IMS. Details of the IMS and CDA are described elsewhere \cite{Vandana_2006,Sumit_2022_CDA}; only a brief description of the experimental scheme is provided here.

The projectile beam delivered from the EBIS overlaps with an effusive CO$_2$ gas jet target at the center of IMS in a cross-beam geometry. The recoil and fragment ions produced in the collision are guided to a ion detector by a uniform extraction field of 60 V/cm applied perpendicular to the direction of both projectile beam and gas jet. The ion detector is a pair of 80 mm diameter microchannel plates in a chevron configuration coupled with a delay line detector to retrieve the two dimensional position ($x,y$) of the particle hitting the detector. The same electric field guides the electrons ejected in the collision to a channeltron detector mounted opposite to ion detector.

The present experiments involved Ar$^{q+}$ ($4\le q\le 16$) impact on CO$_2$ at velocities close to 0.3~a.u.: $v=0.27$~a.u.\ for $q=4$ and $v=0.31$~a.u.\ for $q \ge 6$. Typical beam current of $\approx 10$ pA with background pressure of a few $10^{-7}$ mbar, ensured low accidental coincidences and single collision conditions.

Fragment and recoil ions are detected in coincidence with either ejected electrons or charge-changed projectiles, where detection of either species serves as the trigger for fragment and recoil ion time-of-flight ($t$) measurements. In this work, we used ejected electrons as the trigger. For each collision event, the ($t,x,y$) information of detected particles was recorded in list mode format for offline analysis. Fragment momentum components ($p_x,p_y,p_z$) were reconstructed from the ($t,x,y$) data, enabling determination of fragment kinetic energies and thus the KER. An extraction field of 60~V/cm yielded a three-body KER resolution of approximately 1 eV with $4\pi$ collection efficiency for ions having energies up to 4.8 eV$/q$.

\section{COMPUTATIONAL DETAILS OF THE PECs}

Potential energy curves (PECs) of the ground state of neutral CO$_2$ and several excited states of CO$_2^{3+}$, dissociating into \mbox{O$^+$:\,C$^+$:\,O$^+$}, were calculated using \textit{ab initio} methods within the GAMESS suite of programs \cite{Schmidt_1993}. The reference wavefunction was generated using the complete active space multiconfiguration self-consistent field (CAS-MCSCF) method, followed by second-order multireference M{\o}ller-Plesset (MRMP) perturbation theory to include dynamic electron correlation. Dunning-type correlation-consistent basis sets  augmented with a set of diffuse functions (aug-cc-pVTZ) were employed throughout. The neutral CO$_2$ ground state electronic configuration is $(1\sigma_\text{g})^2$$(1\sigma_\text{u})^2$$(2\sigma_\text{g})^2$$(3\sigma_\text{g})^2$$(2\sigma_\text{u})^2$$(4\sigma_\text{g})^2$$(3\sigma_\text{u})^2$$(1\pi_\text{u})^4$$(1\pi_\text{g})^4$ $^1\Sigma_\text{g}^+$ . In both MCSCF and MRMP calculations, three core orbitals were kept frozen, while all twelve valence orbitals were kept as active orbitals, allowing a balanced treatment of electron correlation. The density convergence criterion was set to $10^{-6}$ to ensure well-converged virtual orbitals. The molecular symmetry was handled within the $D_{2h}$ point group.  PECs were generated over a range of irreducible representations corresponding to different molecular symmetries for doublet, quartet, sextet, and octet spin multiplicities. For all PEC calculations, the geometry of the molecular ion was kept linear, with symmetrical stretching of both C-O bonds from 1.0 to 4.0 Å in increments of 0.02 Å. The calculated bond length for equilibrium geometry of CO$_2$ is  1.17~\AA\ in good agreement with the established value of 1.162~\AA.

\begin{figure}
\centering
\includegraphics[width=\columnwidth]{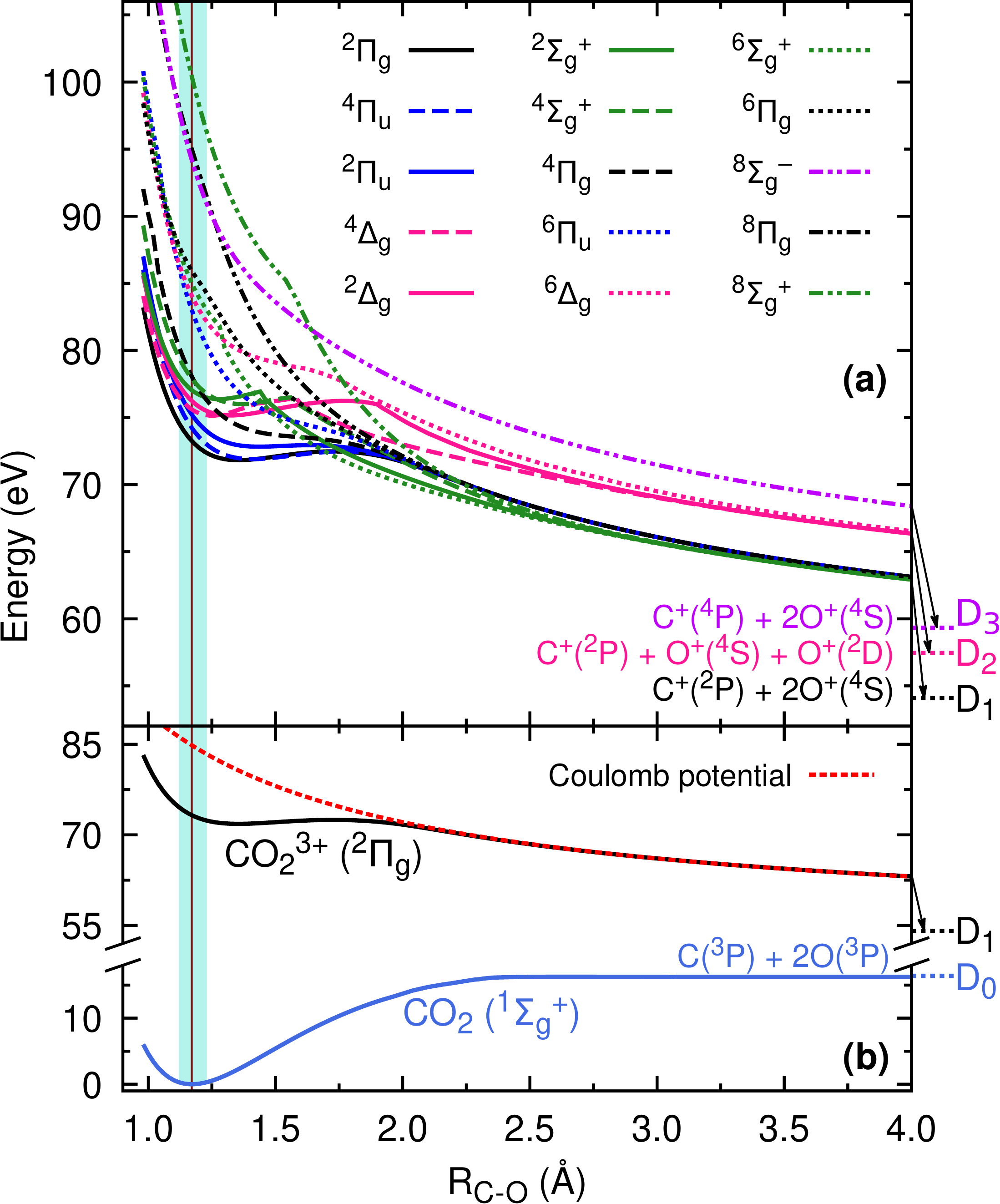}
\caption{(a) Potential energy curves (PECs) of ground and excited states of linear CO$_2^{3+}$, calculated for symmetric stretch of the C--O bonds. D$_1$, D$_2$, and D$_3$ indicate the lowest dissociation limits to which these PECs map asymptotically. (b) PEC of the ground state of CO$_2$ ($^1\Sigma_\text{g}^+$) and CO$_2^{3+}$ ($^2\Pi_\text{g}$). Also shown is the classical Coulomb PEC shifted energetically by D$_1$ (see text). The vertical strip represents the Franck-Condon region. All energies are referenced to the equilibrium geometry of neutral CO$_2$.}
\label{figure1}
\end{figure}

The calculated PECs of CO$_2^{3+}$ corresponding to the lowest adiabatic state of each symmetry and spin multiplicity are shown in Fig.~\ref{figure1}\textcolor{blue}{(a)}, alongside the three lowest dissociation limits to which these states asymptotically map. All energies are reported relative to the equilibrium geometry of neutral CO$_2$ ($^1\Sigma_\text{g}^+$) (cf. Fig.~\ref{figure1}\textcolor{blue}{(b)}). Our calculations agree with those of Wang \emph{et al.} \cite{Wang_2015} with regard to PECs shape and energy referenced to equilibrium geometry of CO$_2$, albeit with a lowering of approximately 2.6 eV for the lowest dissociation limit (D$_1$) compared to their values. To resolve this discrepancy, the PEC of the ground state ($^2\Pi_\text{g}$) of CO$_2^{3+}$ is displayed alongside the classical Coulomb potential energy curve in Fig.~\ref{figure1}\textcolor{blue}{(b)}, shifted energetically by D$_1$ relative to equilibrium geometry of neutral CO$_2$. The classical Coulomb potential energy of the \mbox{O$^+$:\,C$^+$:\,O$^+$} system is given by:
\begin{equation}
\text{V} = 14.4 \left( \frac{2}{R_{\text{C--O}}} + \frac{1}{R_{\text{O--O}}} \right)  + \text{D}_1 \label{eq1}
\end{equation}
where $R_{\text{C--O}}$ and $R_{\text{O--O}}$ are internuclear distances in \AA. The excellent match between the classical Coulomb curve and the PEC of $^2\Pi_\text{g}$ state at large internuclear distances validates the physical accuracy of our calculated dissociation limits.

\section{RESULTS AND DISCUSSION}

The decay of excited states of CO$_2^{3+}$ accessed as a result of the collision can proceed either via a concerted pathway involving nearly simultaneous breaking of both C--O bonds or via a sequential pathway involving the formation of a transient CO$^{2+}$ molecular ion in the first step that subsequently fragments in the second step. To analyze the break-up dynamics of these two pathways, it is essential to distinguish between collision events leading to sequential and concerted fragmentation. Using the native-frames method, we disentangle events associated with each break-up mechanism and analyze their respective dynamics, with a particular emphasis on the KER distributions.

\subsection{Disentangling sequential and concerted break-up}

\begin{figure*}
	\centering
	\includegraphics[width=\textwidth]{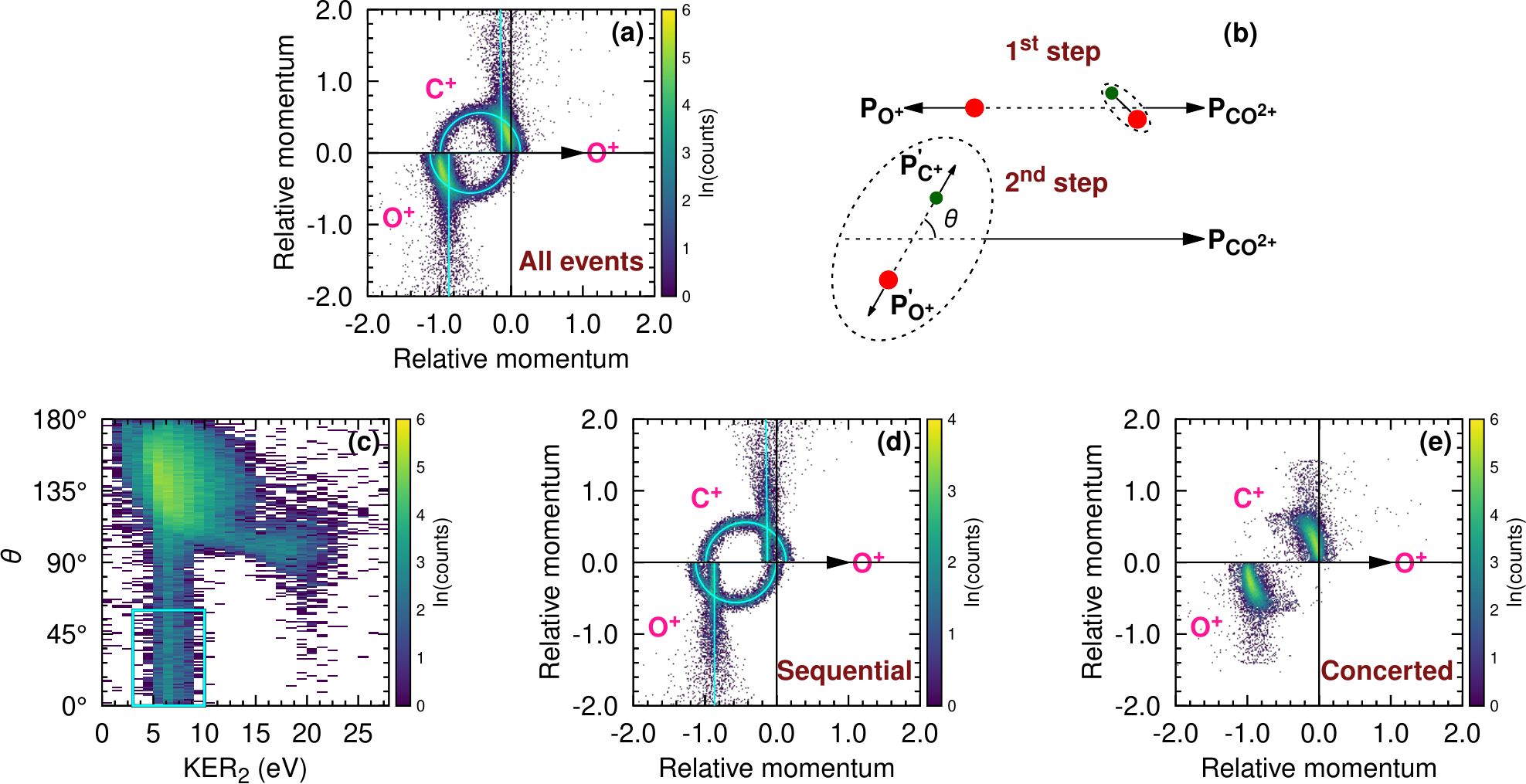}
	\caption{Native-frames analysis of the \mbox{O$^+$:\,C$^+$:\,O$^+$} break-up channel of CO$_2^{3+}$, produced in collisions with 0.31~a.u.\ Ar$^{10+}$ projectiles. (a) Newton diagram for all events. (b) Illustration of the sequential break-up: CO$^{2+}$ formed in the first step subsequently dissociates into C$^+$ and O$^+$ in the second step. Primed quantities denote momentum vectors in the CO$^{2+}$ center-of-mass frame, unprimed in the CO$_2^{3+}$ center-of-mass frame. $\theta$ represents the angle between the directions of the two break-ups. (c) Distribution of all events as a function of the kinetic energy released in the center-of-mass frame of CO$^{2+}$ in second step (KER$_{\text{2}}$) and $\theta$. Events inside the rectangular region are used as a template to reconstruct sequential events outside this region. (d) Newton diagram for sequential events and (e) for concerted events, separated by native-frames analysis. The O$^+$ ion that is detected first is taken as the reference ion in Newton diagrams and is attributed to CO$^{2+}$ break-up.}
	\label{figure2}
\end{figure*}

Newton diagrams and Dalitz plots are standard tools for analyzing sequential and concerted fragmentation mechanisms \cite{Neumann_2010,Sumit_2022,Khan_2015,Wang_2015,Yan_2016,Jana_2011,Bhatt_2012}. In Fig.~\ref{figure2}\textcolor{blue}{(a)} we show the Newton diagram for the \mbox{O$^+$:\,C$^+$:\,O$^+$} break-up channel of CO$_2^{3+}$.  The momentum vector of one of the O$^+$ ions is chosen as the reference direction (along the \textit{x}-axis) and normalized to unity. The momentum vectors of the other O$^+$ ion and the third C$^+$ ion, scaled in magnitude to the reference O$^+$ momentum, are then plotted relative to it in the upper and lower half-planes respectively. Within this representation, sequential break-up events appear as characteristic semi-circular structures, while concerted break-up events form crescent-like patterns. However, the two sets of events have a substantial overlap in this representation. This limits the ability to determine the branching ratios for the two types of break-up pathways. A similar overlap occurs in the Dalitz plot representation also (not shown).

The native-frames method overcomes this limitation \cite{Rajput_2018,Severt_2024}. This method exploits the fact that if the lifetime of the intermediate dication is much longer than its rotational period, then the angle $\theta$ between the first and second break-up steps (see Fig.~\ref{figure2}\textcolor{blue}{(b)}) is expected to have a uniform distribution.

This uniform distribution is readily seen in a 2-D plot of all the events as a function of  $(\text{KER}_{\text{2}},\theta)$ (Fig.~\ref{figure2}\textcolor{blue}{(c)}), where KER$_{\text{2}}$ is the kinetic energy released in the center-of-mass frame of the intermediate CO$^{2+}$ molecular ion. Since either of the two O$^{+}$ ions can separate in the first step, there will be two such distributions. They are analyzed separately.

In the present case sequential break-up events manifest as a uniformly distributed vertical strip centered around KER$_{\text{2}}=$ 6.5~eV. The region where sequential events are unambiguously isolated (marked as a rectangle in Fig.~\ref{figure2}\textcolor{blue}{(c)}) serves as a template to reconstruct the entire sequential-event strip. Details of the reconstruction can be found in \cite{Rajput_2018,Severt_2024}. Such reconstruction permits generation of the distribution of various kinematic variables attributable uniquely to sequential break-up. Subtracting the distribution of sequential events from the total then yields the distribution of concerted events. Separated Newton diagrams for sequential and concerted events are shown in Fig.~\ref{figure2}\textcolor{blue}{(d), (e)}, respectively. The former clearly reveals semi-circular structures associated with sequential events alone, while the latter prominently features crescent-like structures associated with concerted events.

\subsection{Kinetic Energy Release distributions}
\label{section_4b}

\begin{figure*}
	\centering
	\includegraphics[width=0.65\textwidth]{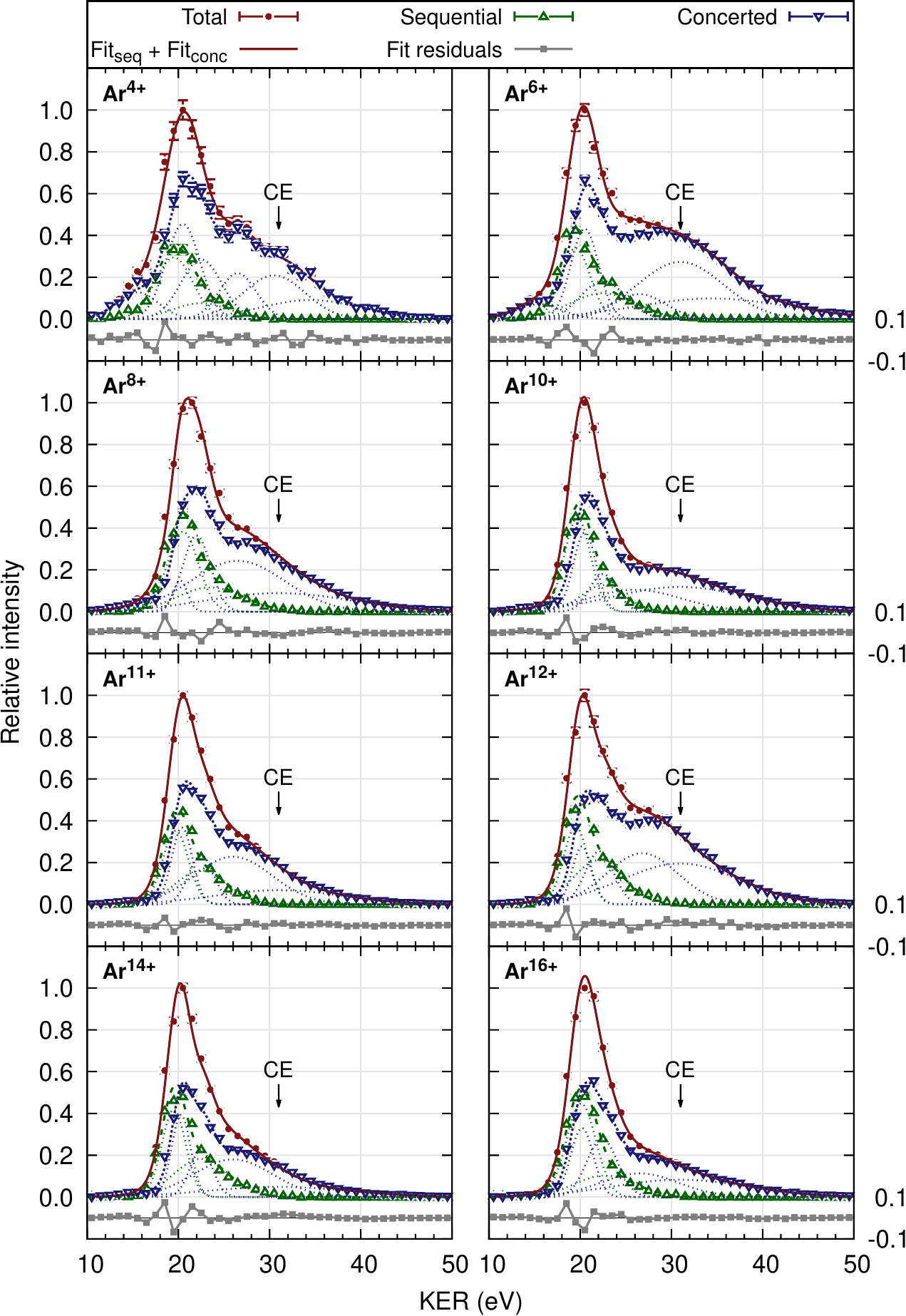}
	\caption{KER distributions for the \mbox{O$^+$:\,C$^+$:\,O$^+$} break-up channel of CO$_2^{3+}$, for various Ar$^{q+}$ projectiles. The total KER distribution is separated into contributions from sequential and concerted pathways using the native-frames method (see text). In each plot, all KER spectra are normalized to the maximum of the total KER distribution. The KER spectra corresponding to sequential and concerted break-up mechanisms are individually fitted to a sum of multiple Gaussians, with the combined fit shown as a solid line. The fit residuals for the combined fit are displayed below each plot. The arrow indicates the KER value expected from the CE model. Error bars on the experimental data represent statistical uncertainties.}
	\label{figure3}
\end{figure*}

We have analyzed the KER distributions for \mbox{O$^+$:\,C$^+$:\,O$^+$} break-up channel of CO$_2^{3+}$, produced in collisions with Ar$^{q+}$ projectiles ($4 \le q \le 16$).  The KER distributions for all events and those for the separated sequential and concerted events are are shown in Fig.~\ref{figure3} for each projectile. The total KER distributions consistently exhibit a pronounced peak at 20.5 eV, accompanied by a broad shoulder above $\sim$25 eV and an extended high-energy tail reaching up to 45–50 eV. The KER distributions for sequential break-up show a prominent peak around 20 eV and remain nearly unchanged with changing projectile charge.  In contrast, KER distributions for concerted break-up show one main peak around 21 eV and a broad shoulder around $\sim$25 eV across all projectiles, with the higher energy feature changing significantly with the projectile charge. The concerted KER distributions exhibit an extended high-energy tail reaching up to 45–50 eV. A distinct low-energy feature near 15.5 eV is also observed for Ar$^{4+}$ projectiles in the concerted break-up, and weakly in the case of Ar$^{6+}$, but it is absent for higher charge states.

The region below $\sim$25~eV KER contains contributions from both sequential and concerted break-up, while the higher-energy part arises predominantly from concerted fragmentation. This observation is consistent with earlier studies that performed similar analyses of the same fragmentation channel \cite{Neumann_2010,Sumit_2022,Khan_2015,Wang_2015}. The overall features of the total KER distribution are also broadly consistent with previous studies involving diverse projectiles \cite{Neumann_2010,Sumit_2022,Khan_2015,Wang_2015,Kumar_2024}, though the relative intensities differ. Notable variations are reported in Refs.~\cite{Yan_2016,Siegmann_2002,Jana_2011}. For instance, Yan \emph{et al.} \cite{Yan_2016} observed a broad KER peak around 34 eV along with a slightly less pronounced peak near 21 eV in collisions with 1.5~a.u.\ Ne$^{4+}$ projectiles. Jana \emph{et al.} \cite{Jana_2011} observed the main KER peak at 26.8 eV for 14.2~a.u.\ Si$^{12+}$ projectiles. Similarly, Siegmann \emph{et al.} \cite{Siegmann_2002}, in collisions with Xe$^{18+}$ (15.4~a.u.), Xe$^{43+}$ (15.4~a.u.), and He$^+$ (1.6~a.u.), reported even higher KER peaks in the 31–35 eV range, with tails extending beyond 70 eV. Moreover, the high-energy structure above 25 eV appears here as a broad shoulder, whereas in electron-impact fragmentation it manifests as a distinct secondary peak around 27 eV \cite{Wang_2015}. In contrast to the multielectron capture mechanism dominant in the present collision study, direct ionization or TI processes are seen to be the principal mechanisms in the previous collision systems at high energies \cite{Yan_2016,Siegmann_2002,Jana_2011}. Furthermore, Yan \emph{et al.} \cite{Yan_2016} found that sequential KER distributions can feature two pronounced peaks at 21 eV and 28.5 eV, which stands in contrast to the single peak near 20 eV observed here for Ar$^{4+}$, despite both projectiles having the same charge. We thus see that differences in fragmentation dynamics are dependent on the principal ionization mechanisms.

The total KER distributions are broadest for Ar$^{6+}$ impact ($\sigma \sim 7.6$ eV) and narrowest for Ar$^{11+}$ ($\sigma \sim 5.4$ eV), consistent with, but somewhat narrower than, previous findings for Ar$^{6+}$ \cite{Kumar_2024}, because of improved high-energy ion collection in the later. The vertical arrow in Fig.~\ref{figure3} marks the KER expected from a classical Coulomb explosion (CE) model. It is clear that the CE model overestimates the observed KER peak and is expected owing to the complete neglect of target electronic structure in this simple model.

To identify the electronic states participating in the types of break-up, the observed features in the sequential and concerted KER distributions are matched with modal KER values calculated from the PECs of CO$_2^{3+}$. These theoretical modal KER values, which are summarized in Table~\ref{table1}, have been calculated under the assumption of Franck-Condon transitions, consistent with the short collision timescales ($\approx$ 1 fs). Both types of KER distributions were fitted with sums of multiple Gaussian distributions with amplitude and width as free parameters for the fit.  Centroids were determined based on distinct spectral features. The number of Gaussians included differed for $q \leq 6$ and $q > 6$, since more features were observed for $q \leq 6$. The centroids and the corresponding most probable electronic states are tabulated in Table~\ref{table2}. In assigning the states, we have included all three dissociation limits listed in Table~\ref{table1}.

Broadly, the same set of electronic states participate across all projectiles. The main peak in the total KER distribution has contributions from both sequential and concerted break-up mechanisms (cf. Fig.~\ref{figure3}), predominantly via low-lying $^{2,4}\Pi_\text{g,u}$ and $^{2,4}\Sigma_\text{g}^+$ states; the $^{2}\Delta_\text{g}$ state may additionally feed this feature through sequential fragmentation. In contrast, the features near 27 eV and 31 eV in the high-KER shoulder are attributed primarily to concerted break-up of high-lying $^{6}\Pi_\text{g,u}$, $^{6}\Delta_\text{g}$, $^{6}\Sigma_\text{g}^+$ and $^{6}\Sigma_\text{g}^+$, $^{6}\Pi_\text{g}$ states respectively. Finally, the extended high-KER tail (beyond $\sim$40 eV) additionally may involve contributions from the $^{8}\Pi_\text{g}$ and $^{8}\Sigma_\text{g}^+$ states.

The pronounced low-energy feature near 15.5~eV observed in the total KER distribution for Ar$^{4+}$ and, less prominently, for Ar$^{6+}$ warrants special attention. This feature typically arises predominantly via sequential break-up in electron impact \cite{Wang_2015} and proton impact \cite{Sumit_2022}. However, in the present case, it primarily results from concerted break-up (cf. Fig.~\ref{figure3}). Wang \emph{et al.} \cite{Wang_2015} attributed this feature to fragmentation via $^2\Pi$ and $^4\Pi$ states, reasoning that the barriers in these states make symmetric stretching of both C–O bonds unlikely.  Instead, asymmetric stretch leads to sequential fragmentation. Here, this feature can arise from fragmentation of the low-lying $^2\Pi_\text{g}$ and $^4\Pi_\text{u}$ states correlating with the D$_2$ dissociation limit or from $^2\Pi_\text{g}$ and $^{2,4}\Pi_\text{u}$ states correlating to the D$_3$ dissociation limit. As shown in Fig.~\ref{figure1}, the states $^2\Pi_\text{g}$, $^2\Pi_\text{u}$, and $^4\Pi_\text{u}$ have barriers at 1.72, 1.64, and 1.76 {\AA}, respectively, but the barriers are lower than the energy at the vertical transition point, allowing these states to also contribute to concerted break-up. It is unlikely that excitation to the same point of the upper PEC would favor concerted break-up in one case and sequential break-up in the other. Hence, in electron \cite{Wang_2015} and proton impact \cite{Sumit_2022}, this feature must arise from different electronic state(s) that strongly favor sequential decay. The $^2\Delta_\text{g}$ and $^4\Delta_\text{g}$ states correlating to the D$_3$ dissociation limit are also possible contributors to this feature; however, they exhibit barriers at 1.78~Å and 1.58~Å, respectively, that lie above the vertical transition point, thereby strongly favoring asymmetric stretching and, consequently, sequential break-up. Therefore, in the present regime, where multielectron capture dominates, the KER feature around 15.5 eV results mainly from concerted break-up involving the lowest three $^2\Pi_\text{g}$ and $^{2,4}\Pi_\text{u}$ states, whereas in electron and proton impact, where direct ionization or TI dominate, this feature predominantly arises from the $^{2,4}\Delta_\text{g}$ states.

\subsection{Influence of projectile charge and electronic structure}
\label{section_4c}

\begin{figure}
	\centering
	\includegraphics[width=\columnwidth]{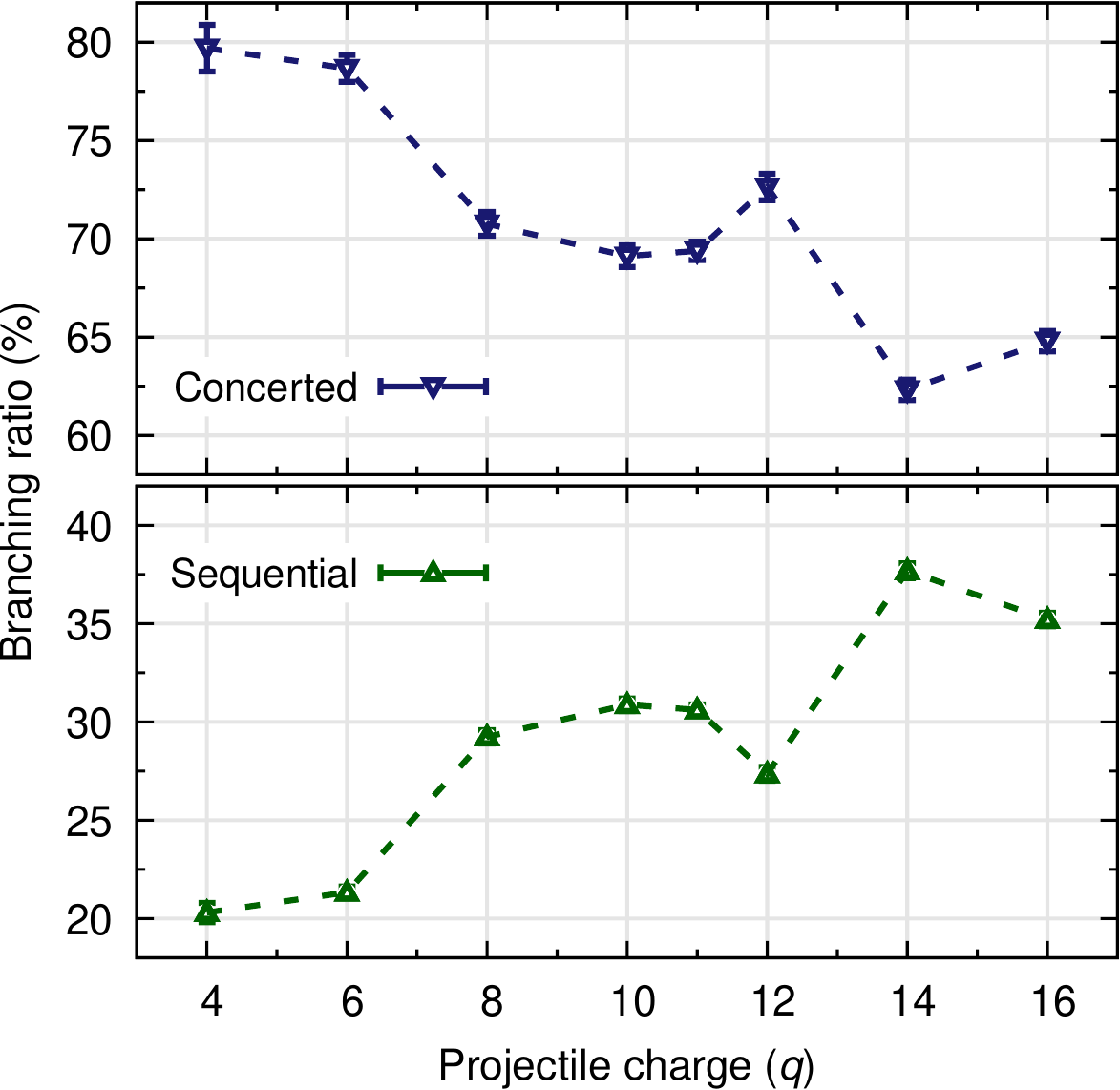}
	\caption{Branching ratios for concerted and sequential break-up as a function of projectile charge $q$. The dashed lines are intended to guide the eye. Error bars on the experimental data represent statistical uncertainties.}
	\label{figure4}
\end{figure}

We now examine the relative probabilities of sequential and concerted break-ups as a function of projectile charge $q$. The native-frames method permits reliable determination of the branching ratios, defined as the ratio of the area under the KER distribution of a given break-up pathway to that of the total KER distribution. Because of limited collection efficiency for high-energy ions, the high-KER region is underrepresented and consequently the branching ratios derived from the KER distributions will be biased in favor of sequential events, (since they do not have a significant high KER contribution). Branching ratios for both break-up pathways as a function of $q$ are presented in Figure~\ref{figure4}. The ratios exhibit a pronounced variation of about 17\% over the present range of $q$. By and large, the fraction of concerted events decreases steadily with increasing $q$ (while that for sequential events increases). Since the sequential mechanism contributes primarily to the low-energy region of the KER distribution, the gradual rise of the sequential branching ratio with increasing $q$ reflects a growing propensity to populate low-lying electronic states. This trend is by-and-large consistent with the predictions of the classical Bohr-Lindhard \cite{Bohr_1954} as well as the Extended Classical Over-the-Barrier (ECOB) model \cite{Niehaus_1986} that the capture radius increases with increasing $q$ and thereby the collision becomes progressively gentler. However, there are deviations from the general trend in case of Ar$^{12+}$ and Ar$^{16+}$. Both models ignore the electronic structure of the projectile, treating it as a bare charge. It is therefore of interest to see how projectiles with different electronic configurations, but the same charge, alter the collision dynamics.

\begin{figure}
	\centering
	\includegraphics[width=\columnwidth]{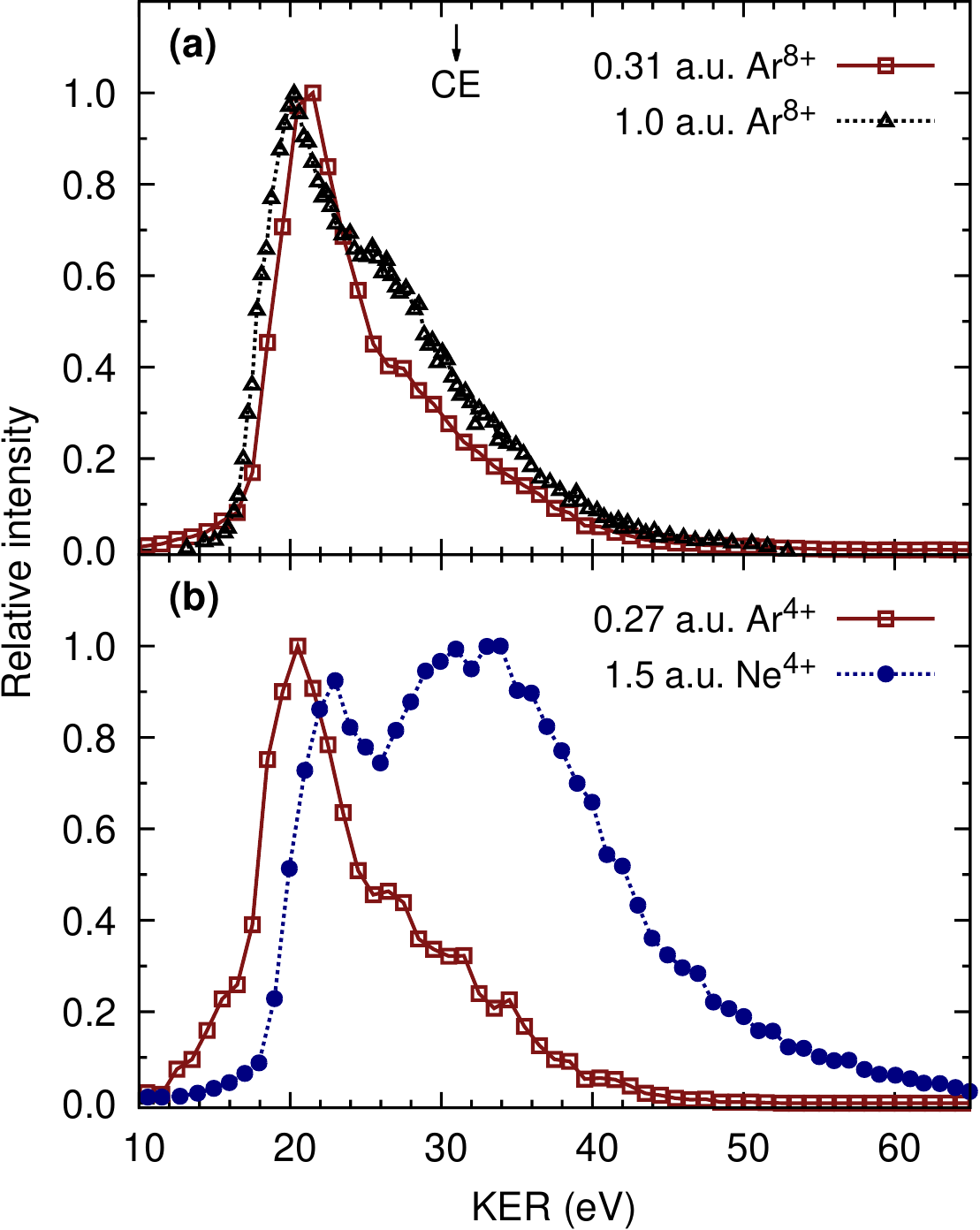}
	\caption{KER distributions for the \mbox{O$^+$:\,C$^+$:\,O$^+$} break-up channel of CO$_2^{3+}$, resulting from collisions with 0.27~a.u.\ Ar$^{4+}$, 0.31~a.u.\ Ar$^{8+}$ (present data), 1.0~a.u.\ Ar$^{8+}$ projectiles \cite{Khan_2015} and 1.5~a.u.\ Ne$^{4+}$ \cite{Yan_2016} projectiles. Each KER distribution is individually normalized to its maximum. The arrow indicates the KER value expected from the CE model.}
	\label{figure5}
\end{figure}

The difference in dynamics can be illustrated by comparing KER distributions for two different projectiles having the same charge. Such data exist in the literature, albeit at different projectile velocities. It is therefore helpful to first compare KER distributions for the same projectile at different velocities and assess the influence of velocity. Figure~\ref{figure5}\textcolor{blue}{(a)} shows the KER distributions for Ar$^{8+}$ projectiles at 0.31~a.u.\ (this work) and 1.0~a.u.\ (Khan \emph{et al.} \cite{Khan_2015}), corresponding predominantly to capture- and TI-dominated regimes, respectively. The two distributions show certain differences, which can be attributed to velocity dependence.

We next compare the KER distributions obtained with two projectiles of identical charge: 0.27~a.u.\ Ar$^{4+}$ (this work) and 1.5~a.u.\ Ne$^{4+}$ (Yan \emph{et al.} \cite{Yan_2016}). These are plotted in Fig.~\ref{figure5}\textcolor{blue}{(b)} and a striking contrast is observed for these equal-charge projectiles. The distribution for 0.27~a.u.\ Ar$^{4+}$ has a clear peak at 20.5~eV, whereas the distribution for 1.5~a.u.\ Ne$^{4+}$ has two peaks: one near 22~eV and the other around 34~eV (close to the value expected from the simple CE model). Based on the data shown in Fig.~\ref{figure5}\textcolor{blue}{(a)}, the pronounced difference in Fig.~\ref{figure5}\textcolor{blue}{(b)} cannot be explained as a consequence of the difference between capture and TI processes. Note that the two ions share the same nominal valence electronic configuration ([Ne]3s$^2$3p$^2$ and [He]2s$^2$2p$^2$, respectively) but the binding energies of their valence shells differ markedly. Thus, while velocity dependence cannot be ignored, the striking contrast between 0.27~a.u.\ Ar$^{4+}$ and 1.5~a.u.\ Ne$^{4+}$ impact must be in a large measure due to the difference in valence shell binding energies of the two projectiles. This comparison highlights the critical role of the projectile’s detailed electronic structure in governing the collision dynamics and may explain the departures from the systematic variations observed in the present study.

In the velocity regime explored in this work, a quasi-molecular picture of the collision process is appropriate \cite{M_Barat_1992}. Within this framework, a transient (ArCO$_2$)$^{q+}$ quasi-molecular complex forms briefly during the collision. Electron capture is then visualized as a transition between molecular states of the entrance channel (Ar$^{q+}$ + CO$_2$) and exit channel (Ar$^{(q-3)+*}$ + CO$_2^{3+*}$). The diabatic PECs associated with these channels can cross at certain internuclear distances. When radial or rotational couplings are present, the corresponding adiabatic PECs exhibit avoided crossings near these geometries. Electron capture is thence viewed as a non-adiabatic transition between adiabatic states correlated to the entrance and exit channels in the vicinity of the avoided crossing. The Landau-Zener transition probability governing this process depends sensitively on factors including the slopes of the PECs near the avoided crossing---a property strongly influenced by the electronic structure of the quasi-molecular constituents.

The captured electrons occupy excited states in the charge-reduced projectile ion, characterized by quantum numbers ($n_1l,n_2l',n_3l''$), where $n$ is the principal and $l$ is the orbital angular momentum quantum number. Measurements resolved in $nl$ show that the captured electrons distribute over a range of $n$ and $l$ values with varying probabilities \cite{Zhang_2017,Xu_2021,Azad_2022}. For any given excited state of CO$_2^{3+}$, there may be multiple avoided crossings with the exit channel PECs, corresponding to certain captured electron configurations ($n_1l,n_2l',n_3l''$). The relative population of this excited state then depends on the number of such crossings and the strength of the coupling between the entrance and exit channels around these points.

This quasi-molecular curve-crossing framework may qualitatively explain the deviations from the systematic trends observed in the KER distributions. Since the KER distributions measured here are integrated over all molecular orientations, the effects of target orientation relative to the projectile trajectory are averaged out in the experimental data. From a theoretical perspective, orientation effects can complicate collision dynamics, so theoretical investigations that incorporate molecular orientation averaging alongside detailed electronic structure considerations are necessary to fully capture and understand the collision dynamics of such systems.

\section{CONCLUSION}

Fragmentation dynamics of the \mbox{O$^+$:\,C$^+$:\,O$^+$} break-up channel of CO$_2^{3+}$ produced in collisions with Ar$^{q+}$ projectiles ($4\le q\le16$) have been investigated at impact velocity of 0.27~a.u.\ for Ar$^{4+}$ and 0.31~a.u.\ for Ar$^{q+}$ ($q\ge6$). Using native-frames analysis, the sequential and concerted contributions were disentangled from the total KER distributions. The main peak around 20.5 eV in the KER spectra arises from both sequential and concerted break-up channels, whereas the high-energy region above $\sim$25 eV is primarily from concerted fragmentation. Correlating the experimental distributions with \textit{ab initio} potential energy curves permit identification of the contributing electronic states of CO$_2^{3+}$. A distinct low-energy feature, previously seen for electron and proton impact, is seen prominently for Ar$^{4+}$ and weakly for Ar$^{6+}$ impact and is attributed to concerted break-up of the low-lying $^2\Pi_\text{g}$, $^2\Pi_\text{u}$, and $^4\Pi_\text{u}$ states.  This contrasts with its attribution to sequential break-up in direct ionization or transfer ionization dominated collisions.

The KER distributions for sequential break-up remain nearly unchanged across all projectiles, whereas the concerted distributions show pronounced yet non-systematic variations with projectile charge $q$. Barring some deviations, the branching ratio for concerted break-up decreases with increasing $q$, while that for sequential break-up increases. However, the observed deviations from systematic trends in the KER distributions and from a simple monotonic behavior in the branching ratios cannot be interpreted within the classical models of electron capture. Comparisons with data in the literature for similar charge states at different velocities reveal a striking contrast in the KER distributions: although velocity effects clearly play an important role, they alone cannot account for the observed differences in the KER spectra. Our findings suggest that it is not charge state of the projectile alone that governs the collision dynamics; the structure of the valence shells of the projectile must also be taken into account. A simple quasi-molecular framework appears to support this point. Further theoretical work within the quasi-molecular framework involving multi-channel couplings is vital for furthering this understanding.

\acknowledgments
The authors thank the Dept.\ of Science and Technology, Science and Engineering Research Board (India) for generous funding via grant No. 30116294, which enabled the setting up of the EBIS/A ion source. They would also like to acknowledge technical help from Dreebit GmbH and assistance from the technical staff at IISER Pune in setting up and running the machine. A. S. acknowledges the award of fellowship by CSIR and thanks Dr. Arnab Sen for valuable help related to GAMESS.

\appendix

\renewcommand{\thetable}{A\arabic{table}}
\setcounter{table}{0}

\section{}

\textit{Ab initio} PECs for various electronic states of CO$_2^{3+}$ have been computed and correlated with the KER data to identify the contributing states. The theoretical modal KER values derived from these PECs, assuming Franck–Condon transitions, are summarized in Table~\ref{table1}. The centroids of the Gaussians used in fitting the sequential and concerted KER distributions (see Section~\ref{section_4b}) and the corresponding most probable electronic states are tabulated in Table~\ref{table2}. In assigning these states, all three dissociation limits listed in Table~\ref{table1} were included. 

\begin{table}
	\centering
	\caption{Calculated modal KER values for various molecular electronic states of CO$_2^{3+}$ fragmenting into \mbox{O$^+$:\,C$^+$:\,O$^+$}. For each electronic state, KER values are provided for the three lowest dissociation limits: D$_1$ (C$^+$($^2$P) + 2 O$^+$($^4$S)), D$_2$ (C$^+$($^2$P) + O$^+$($^4$S) + O$^+$($^2$D)), and D$_3$ (C$^+$($^4$P) + 2 O$^+$($^4$S)). Values corresponding to fragmentation while maintaining linear geometry of the molecular ion and symmetric stretch of both C-O bonds are indicated by the superscript (L).}
	\begin{tabularx}{\columnwidth}{@{}  >{\raggedright\arraybackslash}X *{3}{>{\centering\arraybackslash}X} @{}}
		\toprule
		\multirow{2}{*}{State} & \multicolumn{3}{c}{KER (eV)} \\
		\cmidrule(){2-4}
		& D$_1$ & D$_2$ & D$_3$ \\
		\midrule
		$^2\Pi_\text{g}$ & ~~~$^\text{}$19.1$^{\text{(L)}}$ & 15.8 & 13.9 \\
		$^4\Pi_\text{u}$ & ~~~$^\text{}$20.1$^{\text{(L)}}$ & 16.8 & 14.9 \\
		$^2\Pi_\text{u}$ & ~~~$^\text{}$21.1$^{\text{(L)}}$ & 17.8 & 15.9 \\
		$^4\Delta_\text{g}$ & 21.5 & ~~~$^\text{}$18.1$^{\text{(L)}}$ & 16.3 \\
		$^2\Delta_\text{g}$ & 22.1 & ~~~$^\text{}$18.8$^{\text{(L)}}$ & 16.9 \\
		$^2\Sigma_\text{g}^+$ & ~~~$^\text{}$22.9$^{\text{(L)}}$ & 19.6 & 17.7 \\
		$^4\Sigma_\text{g}^+$ & ~~~$^\text{}$23.8$^{\text{(L)}}$ & 20.4 & 18.6 \\
		$^4\Pi_\text{g}$ & ~~~$^\text{}$23.9$^{\text{(L)}}$ & 20.6 & 18.7 \\
		$^6\Pi_\text{u}$ & ~~~$^\text{}$28.8$^{\text{(L)}}$ & 25.5 & 23.7 \\
		$^6\Delta_\text{g}$ & 29.9 & ~~~$^\text{}$26.6$^{\text{(L)}}$ & 24.7 \\
		$^6\Sigma_\text{g}^+$ & ~~~$^\text{}$30.9$^{\text{(L)}}$ & 27.6 & 25.7 \\
		$^6\Pi_\text{g}$ & ~~~$^\text{}$31.8$^{\text{(L)}}$ & 28.5 & 26.6 \\
		$^8\Sigma_\text{g}^-$ & 40.0 & 36.7 & ~~~$^\text{}$34.8$^{\text{(L)}}$ \\
		$^8\Pi_\text{g}$ & ~~~$^\text{}$40.8$^{\text{(L)}}$ & 37.5 & 35.6 \\
		$^8\Sigma_\text{g}^+$ & ~~~$^\text{}$46.3$^{\text{(L)}}$ & 42.9 & 41.1 \\
		\bottomrule
	\end{tabularx}
	\label{table1}
\end{table}

\begin{table*}
	\centering
	\caption{Centroids of the Gaussians employed in fitting the sequential and concerted KER distributions, together with the most probable electronic states of CO$_2^{3+}$, for all projectiles.}
	\setlength{\tabcolsep}{1pt} 
	\begin{tabular*}{\textwidth}{@{\extracolsep{\fill}} l *{8}{c} @{}}
		\toprule
		Projectile & \multicolumn{2}{c}{Sequential} & \multicolumn{6}{c}{Concerted} \\
		\cmidrule(){1-1} \cmidrule(){2-3} \cmidrule(){4-9}
		Ar$^{4+}$ & 19.5 & 22.5 & 15.5 & 20.5 & 22.5 & 26.5 & 30.5 & 34.5 \\		
		& $^{2,4}\Pi_\text{g}$,$^4\Pi_\text{u}$,$^{2}\Delta_\text{g}$,$^{2,4}\Sigma_\text{g}^+$ & $^{2,4}\Delta_\text{g}$,$^{2}\Sigma_\text{g}^+$ & $^2\Pi_\text{g}$,$^{2,4}\Pi_\text{u}$ & $^{2,4}\Pi_\text{u}$,$^{2,4}\Sigma_\text{g}^+$,$^{4}\Pi_\text{g}$ & $^{2,4}\Delta_\text{g}$,$^{2}\Sigma_\text{g}^+$ & $^{6}\Pi_\text{g,u}$,$^{6}\Delta_\text{g}$,$^{6}\Sigma_\text{g}^+$ & $^{6}\Delta_\text{g}$,$^{6}\Sigma_\text{g}^+$ & $^8\Sigma_\text{g}^-$ \\
		
		& \\
		Ar$^{6+}$ & 19.5 & 22.5 & 15.5 & 20.5 & 22.5 & 26.5 & 31.0 & 34.5 \\
		& $^{2,4}\Pi_\text{g}$,$^4\Pi_\text{u}$,$^{2}\Delta_\text{g}$,$^{2,4}\Sigma_\text{g}^+$ & $^{2,4}\Delta_\text{g}$,$^{2}\Sigma_\text{g}^+$ & $^2\Pi_\text{g}$,$^{2,4}\Pi_\text{u}$ & $^{2,4}\Pi_\text{u}$,$^{2,4}\Sigma_\text{g}^+$,$^{4}\Pi_\text{g}$ & $^{2,4}\Delta_\text{g}$,$^{2}\Sigma_\text{g}^+$ & $^{6}\Pi_\text{g,u}$,$^{6}\Delta_\text{g}$,$^{6}\Sigma_\text{g}^+$ & $^{6}\Sigma_\text{g}^+$,$^{6}\Pi_\text{g}$ & $^8\Sigma_\text{g}^-$ \\
		
		& \\
		Ar$^{8+}$ & 20.5 & 23.5 & & 20.2 & 22.0 & 26.5 & 31.0 & \\
		& $^{2,4}\Pi_\text{u}$,$^{4}\Delta_\text{g}$,$^{4}\Sigma_\text{g}^+$,$^{4}\Pi_\text{g}$ & $^{2,4}\Sigma_\text{g}^+$,$^{4}\Pi_\text{g}$,$^{6}\Pi_\text{u}$ & & $^{2,4}\Pi_\text{u}$,$^{2,4}\Sigma_\text{g}^+$,$^{4}\Pi_\text{g}$ & $^{2}\Pi_\text{u}$,$^{2,4}\Delta_\text{g}$,$^{2}\Sigma_\text{g}^+$ & $^{6}\Pi_\text{g,u}$,$^{6}\Delta_\text{g}$,$^{6}\Sigma_\text{g}^+$ & $^{6}\Sigma_\text{g}^+$,$^{6}\Pi_\text{g}$ & \\
		
		& \\
		Ar$^{10+}$ & 19.7 & 22.5 & & 20.5 & 22.5 & 26.0 & 31.0 & \\
		& $^{2,4}\Pi_\text{g}$,$^4\Pi_\text{u}$,$^{2}\Delta_\text{g}$,$^{2,4}\Sigma_\text{g}^+$ & $^{2,4}\Delta_\text{g}$,$^{2}\Sigma_\text{g}^+$ & & $^{2,4}\Pi_\text{u}$,$^{2,4}\Sigma_\text{g}^+$,$^{4}\Pi_\text{g}$ & $^{2,4}\Delta_\text{g}$,$^{2}\Sigma_\text{g}^+$ & $^{6}\Pi_\text{g,u}$,$^{6}\Delta_\text{g}$,$^{6}\Sigma_\text{g}^+$ & $^{6}\Sigma_\text{g}^+$,$^{6}\Pi_\text{g}$ & \\
		
		& \\
		Ar$^{11+}$ & 19.7 & 22.5 & & 20.4 & 22.5 & 26.0 & 31.0 & \\
		& $^{2,4}\Pi_\text{g}$,$^4\Pi_\text{u}$,$^{2}\Delta_\text{g}$,$^{2,4}\Sigma_\text{g}^+$ & $^{2,4}\Delta_\text{g}$,$^{2}\Sigma_\text{g}^+$ & & $^{2,4}\Pi_\text{u}$,$^{2,4}\Sigma_\text{g}^+$,$^{4}\Pi_\text{g}$ & $^{2,4}\Delta_\text{g}$,$^{2}\Sigma_\text{g}^+$ & $^{6}\Pi_\text{g,u}$,$^{6}\Delta_\text{g}$,$^{6}\Sigma_\text{g}^+$ & $^{6}\Sigma_\text{g}^+$,$^{6}\Pi_\text{g}$ & \\
		
		& \\
		Ar$^{12+}$ & 19.5 & 22.0 & & 20.2 & 22.1 & 26.8 & 31.0 & \\
		& $^{2,4}\Pi_\text{g}$,$^4\Pi_\text{u}$,$^{2}\Delta_\text{g}$,$^{2,4}\Sigma_\text{g}^+$ & $^{2,4}\Delta_\text{g}$,$^{2}\Sigma_\text{g}^+$ & & $^{2,4}\Pi_\text{u}$,$^{2,4}\Sigma_\text{g}^+$,$^{4}\Pi_\text{g}$ & $^{2,4}\Delta_\text{g}$,$^{2}\Sigma_\text{g}^+$ & $^{6}\Pi_\text{g}$,$^{6}\Delta_\text{g}$,$^{6}\Sigma_\text{g}^+$ & $^{6}\Sigma_\text{g}^+$,$^{6}\Pi_\text{g}$ & \\
		
		& \\
		Ar$^{14+}$ & 19.5 & 22.5 & & 20.3 & 22.5 & 26.0 & 31.0 & \\
		& $^{2,4}\Pi_\text{g}$,$^4\Pi_\text{u}$,$^{2}\Delta_\text{g}$,$^{2,4}\Sigma_\text{g}^+$ & $^{2,4}\Delta_\text{g}$,$^{2}\Sigma_\text{g}^+$ & & $^{2,4}\Pi_\text{u}$,$^{2,4}\Sigma_\text{g}^+$,$^{4}\Pi_\text{g}$ & $^{2,4}\Delta_\text{g}$,$^{2}\Sigma_\text{g}^+$ & $^{6}\Pi_\text{g,u}$,$^{6}\Delta_\text{g}$,$^{6}\Sigma_\text{g}^+$ & $^{6}\Sigma_\text{g}^+$,$^{6}\Pi_\text{g}$ & \\
		
		& \\
		Ar$^{16+}$ & 20.1 & 22.7 & & 20.2 & 22.4 & 26.5 & 31.0 & \\
		& $^{2,4}\Pi_\text{g}$,$^{2,4}\Pi_\text{u}$,$^{2,4}\Sigma_\text{g}^+$ & $^{2}\Delta_\text{g}$,$^{2}\Sigma_\text{g}^+$,$^{6}\Pi_\text{u}$ & & $^{2,4}\Pi_\text{u}$,$^{2,4}\Sigma_\text{g}^+$,$^{4}\Pi_\text{g}$ & $^{2,4}\Delta_\text{g}$,$^{2}\Sigma_\text{g}^+$ & $^{6}\Pi_\text{g,u}$,$^{6}\Delta_\text{g}$,$^{6}\Sigma_\text{g}^+$ & $^{6}\Sigma_\text{g}^+$,$^{6}\Pi_\text{g}$ & \\
		\bottomrule
	\end{tabular*}
	\label{table2}
\end{table*}

%


\end{document}